\documentclass[pra,showpacs,preprintnumbers,twocolumn,amsmath,amssymb]{revtex4}
\usepackage{graphicx}% Include figure files 
\usepackage{color}
\usepackage{amsmath}
\usepackage{epstopdf}
\usepackage{amsbsy}
\usepackage{dcolumn}% Align table columns on decimal point
\usepackage{bm}% bold math

\begin{document}

\title{Inducing spin-dependent tunneling to probe magnetic correlations in optical lattices}

\author{K.\ G.\ L.\ Pedersen$^1$}
\author{B.\ M.\ Andersen$^1$}
\author{O.\ F.\ Sylju\aa sen$^2$}
\author{G.\ M.\ Bruun$^3$}
\author{A.\ S.\ S\o rensen$^1$}
\affiliation{
$^1$Niels Bohr Institute, University of Copenhagen, DK-2100 Copenhagen \O, Denmark\\
$^2$Department of Physics, University of Oslo, P. O. Box 1048 Blindern, N-0316 Oslo, Norway\\
$^3$Department of Physics and Astronomy, Aarhus University, Ny Munkegade, DK-8000 Aarhus C, Denmark
}

\date{\today}

\begin{abstract}
We suggest a simple experimental method for probing antiferromagnetic spin correlations of two-component Fermi gases in optical lattices. The method relies on a spin selective Raman transition to excite atoms of one spin species to their first excited vibrational mode where the tunneling is large. The resulting difference in the tunneling dynamics of the two spin species can then be exploited,  to reveal the spin correlations by measuring the number of doubly occupied lattice sites at a later time. We perform quantum Monte Carlo simulations of the spin system and solve the optical lattice dynamics numerically to show how the timed probe can be used to identify antiferromagnetic spin correlations in optical lattices.
\end{abstract}

\pacs{67.85.d, 73.21.Cd, 74.25.q, 75.50.Ee}

\maketitle

\section{Introduction}
Atoms in optical lattices provide a powerful test bed for probing the properties of quantum systems in periodic potentials. 
Several fundamental results have been reported including the observation of Mott physics~\cite{Mott}, 
fermionic pairing~\cite{Chin}, and the  fermionization of bosons in one dimension~\cite{Paredes}. Promising new research directions 
are emerging with the creation of Dirac points in honeycomb lattices~\cite{Tarruell} and the achievement of  single site resolution for
 probing and manipulating  two dimensional (2D) lattices~\cite{ShersonBakr,Weitenberg}.
A major objective in the study of atoms in optical lattices is to unravel the physics of quantum magnetism,
 and for this purpose it is crucial to be able to measure the spin correlations 
of the atoms efficiently. Several methods have been suggested including Bragg scattering~\cite{Corcovilos}, time-of-flight measurements~\cite{Altman,BSPADS,Andersen}, and 
light polarization experiments~\cite{BADS}. A complication for these methods is, however, that the experimental signal is small,
 and that one has to average over many experimental realizations.

To address this, we recently suggested a superlattice method which yields large macroscopic signatures of the spin correlations and allows for 
the detection of both short- 
and long-range correlations~\cite{pedersen}. It is based on  spin-dependent transport using a spin-selective Raman excitation followed by the application of an additional moving 
superlattice potential. The spin-correlations can then be extracted  by measuring the number of doubly occupied lattice sites.

 In this paper we propose a similar but experimentally simpler method, which replaces the moving superlattice by a precisely timed period of free tunneling dynamics of the atoms followed by a measurement of the double occupancy. We analyze the spin dependence of the tunneling dynamics and show how this can be used to reveal the spin correlations in the initial state. The main result of our paper is the demonstration that  timed measurements of the probability for double occupancy of the lattice sites 
   yields distinct signatures of the magnetic correlations.
  
Importantly, the method we propose here does not require any spin dependence of the lattice potential and only utilizes only the optical lattice potential which is already present. This is a major advantage from an experimental perspective.

\section{Model and Method}
We consider a  gas of fermionic atoms  in an optical lattice with equal populations in two internal states $|\sigma\rangle$ ($\sigma=\uparrow,\downarrow$), which we refer to as spin in the following. 
For strong repulsion and half-filling,  the gas is in the Mott phase at low temperature $T$  which is well described by the Heisenberg model~\cite{Duan}
\begin{equation}
\hat H= J \sum_{\langle l,m\rangle} [ \hat{s}^x_l \hat{s}^x_m +  \hat{s}^y_l \hat{s}^y_m +  (1+\Delta) \hat{s}^z_l \hat{s}^z_m].
\label{Heisenberg}
\end{equation} 
Here, $\mathbf{\hat{s}}_l$ is the spin-$1/2$ operator for atoms at site $l$, and $\left\langle l,m \right\rangle$ denotes neighboring pairs. The interaction strength $J$ can exhibit a small anisotropy, quantified by the anisotropy parameter $\Delta$. 
A major incentive for studying atoms in optical lattices is to observe the phase transition to an antiferromagnetic (AF) ordered state which occurs at $T=T_N\simeq 0.945J$ in 3D for the isotropic ($\Delta = 0$) Heisenberg model~\cite{BSPADS}. To achieve this, one needs a sensitive experimental probe of the spin correlations. 

The key idea of the present paper is to access the spin correlation functions by inducing a difference in the tunneling dynamics of the two different spin states $| \! \uparrow \rangle$ and $| \!\downarrow \rangle$ as shown in the sketch of Figure \ref{fig:timeevolution}(a). To 
this end, we excite all $| \! \uparrow \rangle$  atoms to their first excited vibrational state in the lattice sites while the  $|\!\downarrow \rangle$ atoms remain in the  ground state~\cite{bouchoule}. This can be done experimentally by set of spin selective Raman transitions on the vibrational sideband of an auxiliary state. Then we simply allow the system to evolve for some time and subsequently measure the number of sites which are doubly occupied $N_d$. Since the tunneling to neighboring sites is larger for the first excited state than for the ground state, the $|\!\downarrow \rangle$ atoms essentially remains at  the initial lattice site as shown in Figure \ref{fig:timeevolution}(a). In practice, the number of doubly occupied sites can be counted  by merging two atoms residing on the same lattice site to a molecule by application of a photo-association pulse or via a Feshbach resonance.\cite{Mott,greif} 

In our theoretical modeling the lattice potential is held constant during the procedure and we assume that the interaction is tuned to a sufficiently weak value, for instance by using a Feshbach resonance~\cite{Stewart}, so that the system can be considered non-interacting during the tunneling dynamics. The  three spatial components of the atomic wave function then evolve independently. Assuming  the motion in the $y$- and $z$-directions to be frozen by strong lattice potentials, it is sufficient to consider the dynamics 
along the $x$-direction, where the atoms are trapped by a periodic potential $A(x,t) = A \sin^2(\pi x/d)$ with $d$  the lattice constant. 
With this potential the single particle Bloch eigenenergies $E_k^l$ and corresponding eigenfunctions $|\psi_k^l \rangle$ are indexed by a vibrational state label $l = 0,1,2,\ldots$ and  a Bloch wavevector $k$. The initial  states are taken to be the Wannier functions given by $|W_{j}^n  \rangle = \frac{1}{\sqrt{N}} \sum_k e^{-ikjd} | \psi^n_k \rangle$, which describe a localized state of the $n$'th vibrational mode centered around lattice site $j$~\cite{BrennenJaksch}.

Because the Wannier functions are not eigenfunctions of the non-interacting lattice Hamiltonian, they evolve over time. To describe this we introduce the transition matrix 
$\hat U(t) = e^{-i\hat H t} = \sum_{ljn} U^{(n)}_{lj}(t) \hat a^{\dagger}_{jn} \hat a^{\phantom{\dagger}}_{ln}$, where $\hat a^\dagger_{jn}$ is the creation operator for the Wannier state $|W_{j}^n  \rangle$ and we use units for which  $\hbar=1$.
The time evolution operator $\hat U(t)$ is quadratic in $\hat a$ and $\hat a^\dagger$ and diagonal in the vibrational quantum number $n$ since we assume the system to be non-interacting during the evolution. 
The matrix elements of this transition matrix read
\begin{equation}
U^{(n)}_{lj}(t) = \langle W_{j}^{n} | W_{l}^n(t)  \rangle = \frac{1}{N} \sum_{k} e^{i k (j -l)d} e^{- i E^n_k t},
\label{Ullnm}
\end{equation}
where $ | W_{l}^n(t)  \rangle$ represents the initial Wannier function $| W_{l}^n \rangle$ evolved for a time $t$. We are mainly interested in the probability for a site $i$ to be doubly occupied $P_d(i) = \langle \hat{n}_{i\uparrow} (t) \hat{n}_{i\downarrow} (t) \rangle$.
Our proposed scheme is based on the fact that this probability can be  
 expressed in terms of spin correlation functions of the initial state and the overlap matrix elements $U^{(n)}_{lj}$~\cite{pedersen}. 
 Since the $| \! \downarrow \rangle$ atoms remain in the vibrational ground state $n=0$ while the $|\! \uparrow \rangle$ atoms are transferred to the first excited state $n=1$, 
 we write $U_{ij}^{(1)}=U_{ij}^{\uparrow}$ and $U_{ij}^{(0)}=U_{ij}^{\downarrow}$. Then we  obtain
\begin{align}
	P_d(i) = &\sum_{jl} | U^\uparrow_{ ij} (t) |^2 | U^\downarrow_{ il}(t)|^2 \left( \tfrac{1}{4} - \left\langle \hat s^z_j \hat s^z_l \right\rangle \right) \nonumber 
	\\&
	- \sum_{j\neq l} U^\uparrow_{ ij} U^\uparrow_{ i l} U^\downarrow_{ i l} U^\downarrow_{ ij} \left\langle \hat s^+_j \hat s^-_l \right\rangle.
\label{PdEqn}
\end{align}
In the limit where the $|\!\downarrow\rangle$ atoms do not move $|U^\downarrow_{ il}|^2 \approx \delta_{i,l}$, Eq. (\ref{PdEqn}) reduces to 
\begin{equation}
	P_d(i) = \frac 1 4- \sum_{j} \left| U^\uparrow_{ ij} (t) \right|^2  \left\langle \hat s^z_j \hat s^z_i \right\rangle. 
\label{Pdsimple}
\end{equation}
Equations (\ref{PdEqn}) and (\ref{Pdsimple}) 
  demonstrate the direct link between the observed double occupancy and the spin correlation function $
\langle \hat s^z_j \hat s^z_i \rangle$ describing  AF ordering in the initial state.

\section{Results}
We now present our numerical results. The spin correlation functions $\langle {\bf \hat s}_l \cdot {\bf \hat s}_j \rangle$ for the initial state determined by the 
3D Heisenberg model (\ref{Heisenberg}) are obtained from  full quantum Monte Carlo (QMC) simulations using the stochastic series expansion method \cite{SS} 
with directed-loop updates \cite{SSE}. The subsequent time evolution after the $| \! \uparrow \rangle$  atoms are excited to the first vibrational state is then calculated  
by numerical diagonalization of the non-interacting   Hamiltonian taking the motion to be along a single direction. As an example, we will here present results for 
a 1D lattice with $N  = 32$  sites and a lattice height $A = 10 E_R$ where  $E_R = \hbar^2 \pi^2 / 2 M d^2$ is the recoil energy with $M$ the atom mass. The numerical 
diagonalization is performed using a lattice unit cell consisting of 1028 real-space points.

\subsection{Single atom dynamics}
\begin{figure}[t]
\includegraphics[clip=true,width=1\columnwidth]{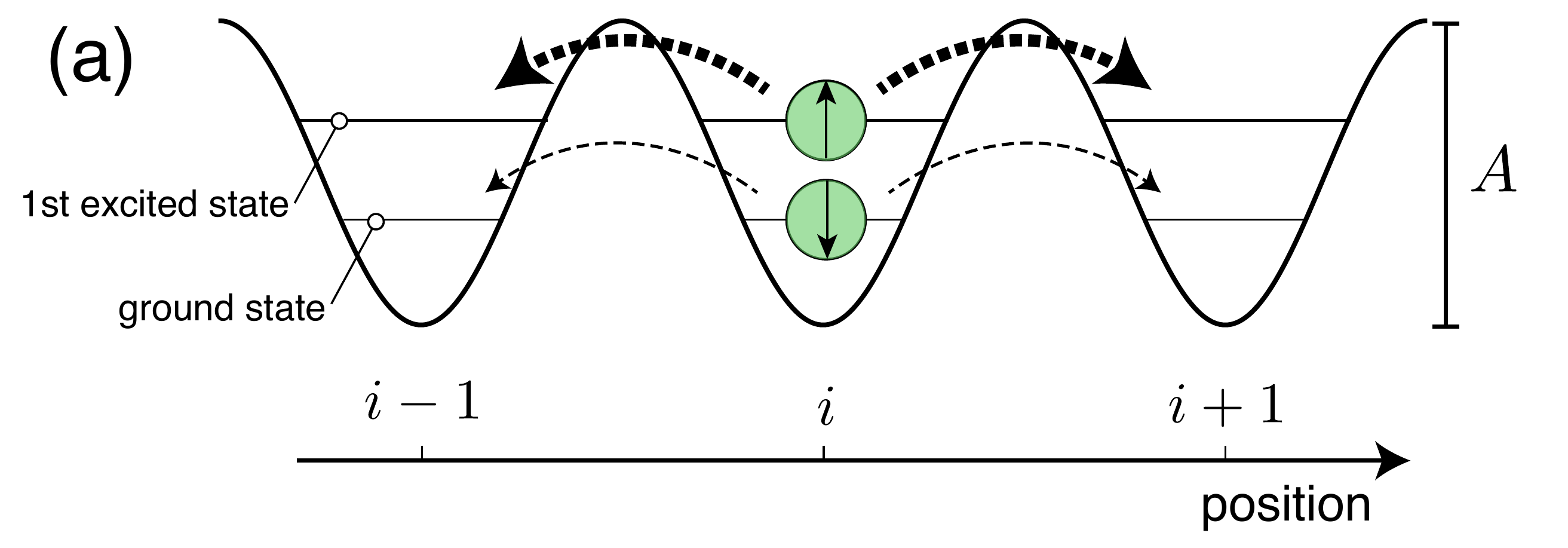}
\includegraphics[clip=true,width=0.98\columnwidth]{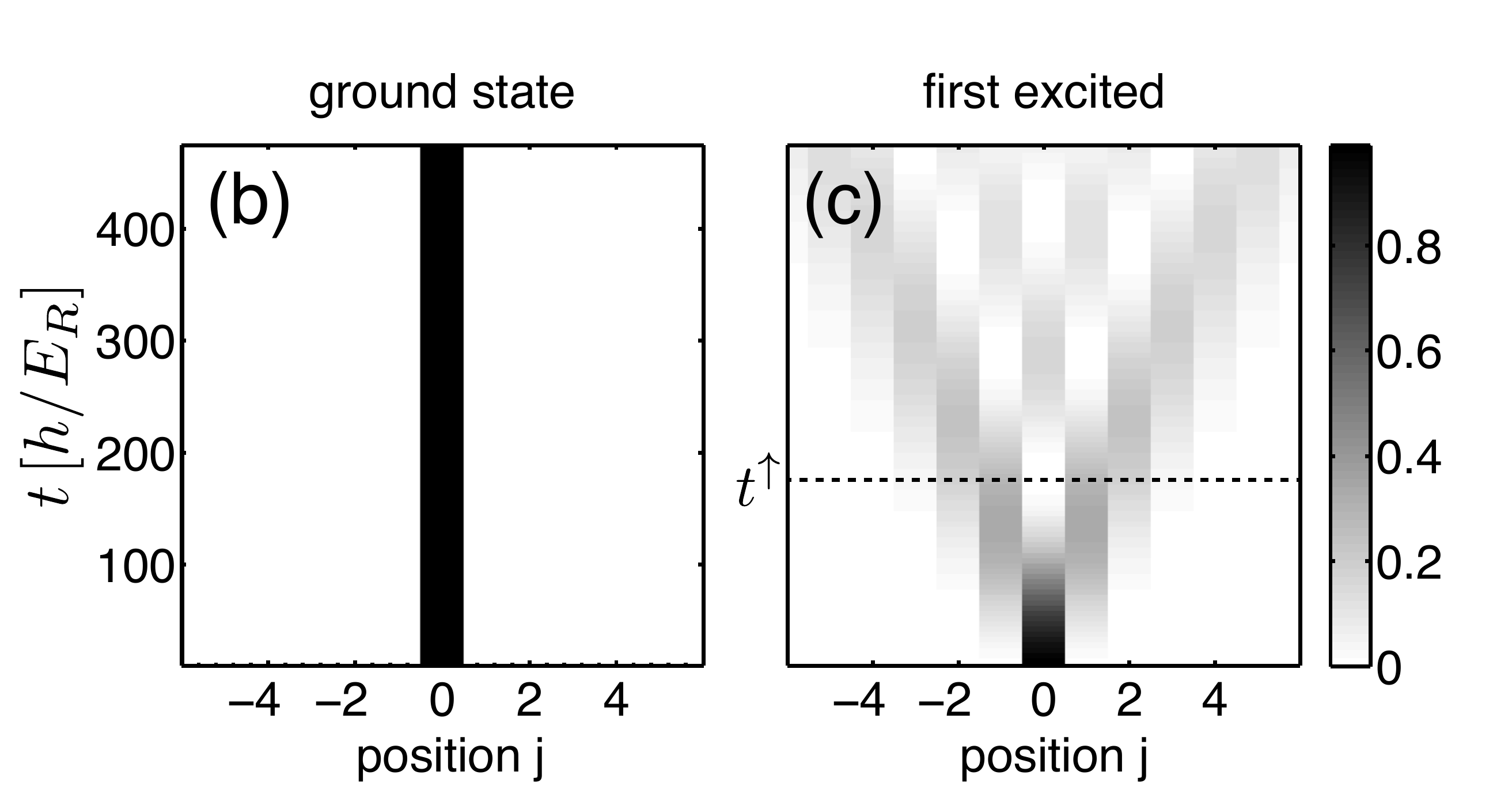}
\caption{(a) A cartoon depicting the tunneling dynamics for the two spin species. (b-c) The actual tunneling dynamics for the two lowest lying vibrational Wannier states of a one-dimensional lattice with lattice height $A = 10 E_R$. We show the norm square of the wavefunction projected onto Wannier states centered around each lattice site. There is a clear difference in the tunneling dynamics between (b) the almost stationary vibrational ground state 
($|\!\downarrow\rangle$ atom) 
and (c) the first excited state ($|\!\uparrow\rangle$ atom) 
 which tunnels  to its neighboring sites. The dashed line indicates the first minimum of the probability for the $|\!\uparrow\rangle$ atom to remain 
 at its initial position.  
 }
\label{fig:timeevolution}
\end{figure}

\begin{figure}
\begin{minipage}{.49\columnwidth}
\includegraphics[clip=true,width=.99\columnwidth]{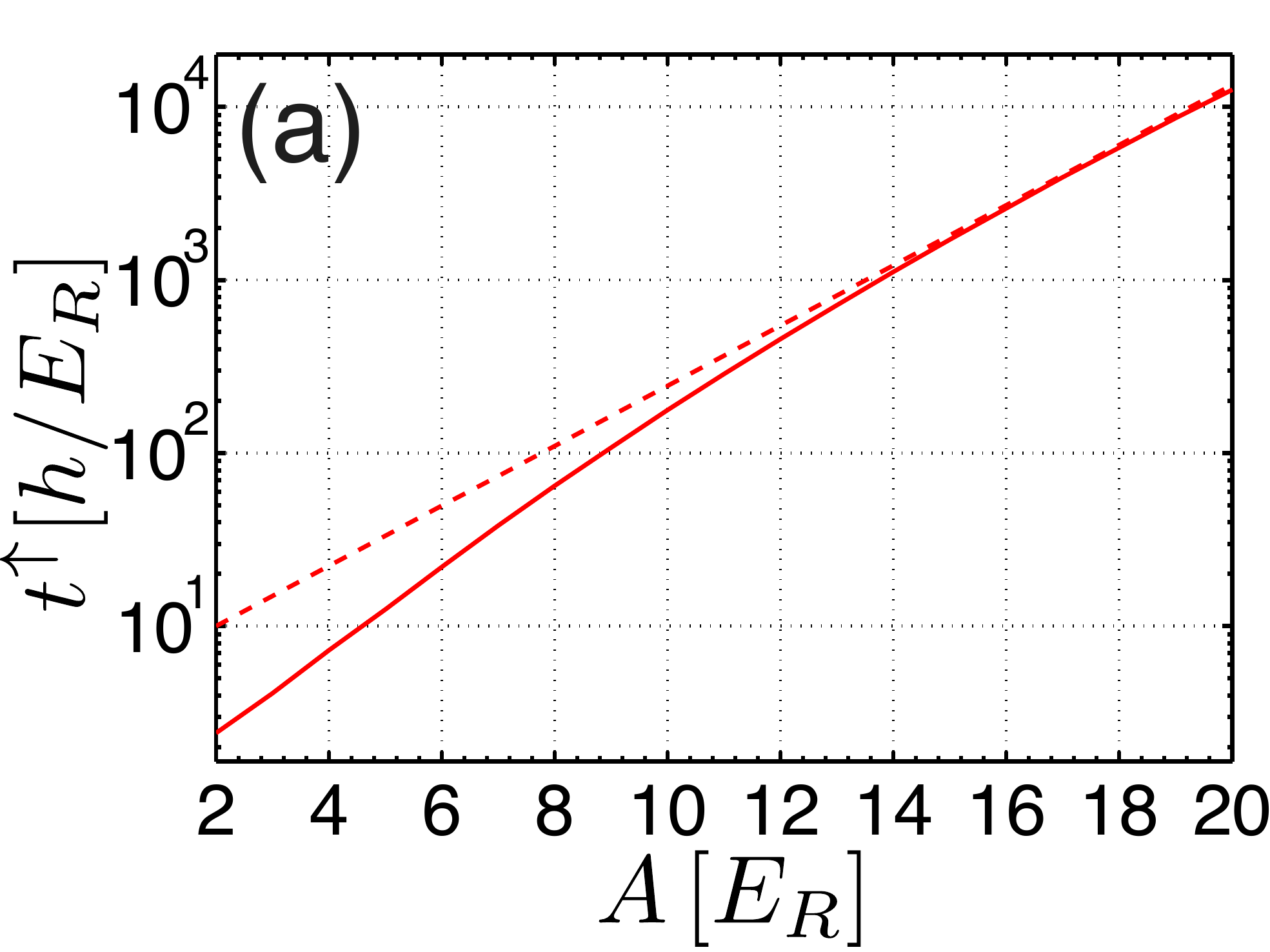}
\end{minipage}
\begin{minipage}{.49\columnwidth}
\includegraphics[clip=true,width=.99\columnwidth]{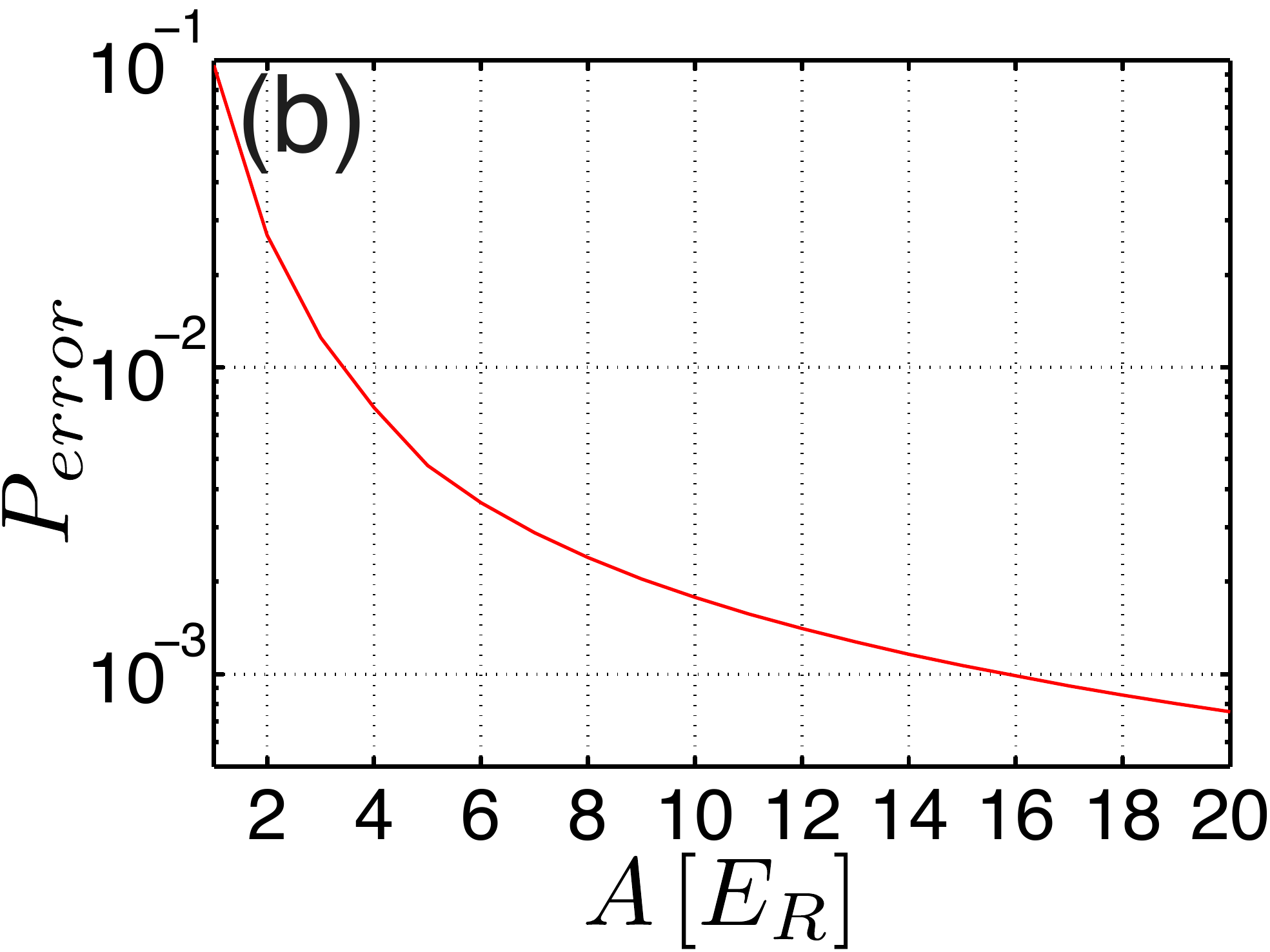}
\end{minipage}
\caption{(Color online) (a) the tunneling time versus lattice depth $A$ of the first excited vibrational  state which sets the time scale 
of the probe. (b) the error probability as a function of lattice depth $A$.}
\label{fig:times}
\end{figure}

We first consider the time evolution of a single atom. Figurew \ref{fig:timeevolution}(b) and \ref{fig:timeevolution}(c) depicts the probability $|U^\sigma_{ 0j} (t) |^2$ of a $|\sigma\rangle$ atom which starts at $t=0$ at site $i=0$,  
to be at site $j$ at time $t$. We see that for the parameters chosen,  the $|\!\downarrow\rangle$ atoms do not move but remain at their initial position (Fig.~\ref{fig:timeevolution}(b)) whereas 
the $|\!\uparrow\rangle$ atoms move (Fig.~\ref{fig:timeevolution}(c)). This makes this tunneling dynamics well suited for measuring AF ordering.
Note that at certain times the wavefunction of the $|\!\uparrow\rangle$ atoms mostly samples either  even or odd lattice sites in a way similar to a  random walk in 1D. 
%%%%%%%%%%%%%%%%%%%%%%%%%%%%%%%%%%%%%%%%%%%%%%%%%%%%%%%%%%%%%%%%%%%
%%%%%%%%%%%%%%%%%%%%%%%%%%%%%%%%%%%%%%%%%%%%%%%%%%%%%%%%%%%%%%%%%%%
 In Fig.~\ref{fig:times}(b) we plot the 
tunneling time of the $|\! \uparrow \rangle$ atoms, $t^\uparrow$, defined as the first zero of $\partial_t |U^\uparrow_{ 00} (t) |^2$, as a function of the lattice depth. According to the WKB approximation this 
 time depends exponentially on the lattice height, and we indeed find $t^\uparrow \propto \exp(0.4 A/E_R)$ for deep lattices.
Thus, by adjusting the lattice depth one can tune the time scale of the probe. It is, however, important that the lattice depth is sufficiently deep so that only the  $|\!\uparrow\rangle$ atoms move. This is 
 illustrated in Fig.~\ref{fig:times} (right) where we plot $P_{\rm error} = 1-|U_{ii}^{(1)}(t^\uparrow)|^2$, which represents the probability that a $|\! \downarrow \rangle$ has moved after the $|\! \uparrow\rangle$ tunneling time $t^\uparrow$. As expected, the error decreases rapidly with increasing lattice depth. 
 %%%%%%%%%%%%%%%%%%%%%%%%%%%%%%%%%%%%%%%%%%%%%%%%%%%%%%%%%%%%%%%%%%%
%%%%%%%%%%%%%%%%%%%%%%%%%%%%%%%%%%%%%%%%%%%%%%%%%%%%%%%%%%%%%%%%%%%

\subsection{Many-body dynamics and AF ordering}
\begin{figure}[b]
\includegraphics[clip=true,width=1\columnwidth]{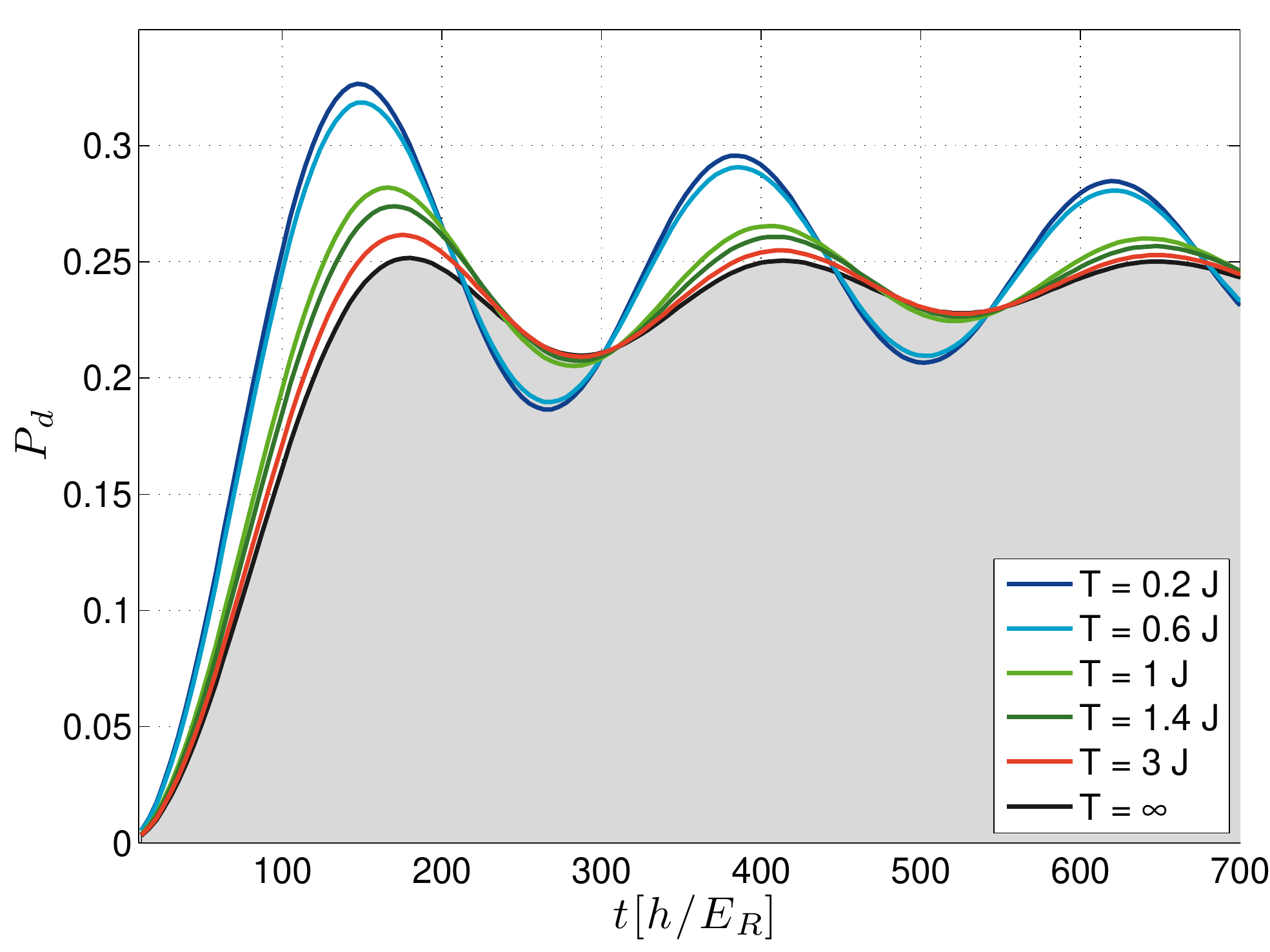}
\caption{(Color online) The double occupancy probability $P_d$ as a function time $t$ at different temperatures $T$ for the slightly anisotropic Heisenberg system ($\Delta = 1.01$) and a lattice depth $A = 10 E_R$.
}
\label{fig:Pd2}
\end{figure}

We now analyze the evolution of the double occupancy, $P_d$ starting from a correlated initial state determined by the 3D Heisenberg model.
This is shown in Fig.~\ref{fig:Pd2}.
%%%%%%%%%%%%%%%%%%%%%%%%%%%%%%%%%%%%%%%%%%%%%%%%%%%%%%%%%%%%%%%%%%%
%%%%%%%%%%%%%%%%%%%%%%%%%%%%%%%%%%%%%%%%%%%%%%%%%%%%%%%%%%%%%%%%%%%

We focus first on the non-interacting case corresponding to $T = \infty$ which is plotted as the solid line forming the boundary to the grey area. For this case, the spins are uncorrelated and 
$\left\langle \hat s^z_m \hat s^z_n \right\rangle = \tfrac{1}{4} \delta_{nm}$. Since the vibrational ground state is essentially immobile 
$|U_{ij}^{\downarrow}|^2 \approx \delta_{i,j}$ 
for all the relevant times, we find \mbox{$P_d \approx \tfrac{1}{4} (1- |U^\uparrow_{ ii} (t)|^2)$}. 
Hence for a half-filled non-interacting system, $P_d$ cannot be larger than 
 $1/4$ and will in general be lower because of the non-vanishing probability that the $|\!\uparrow\rangle$ atoms return to their initial states. The oscillations in $P_d$ for uncorrelated atoms are therefore directly linked to the oscillations in $|U^\uparrow_{ ii} (t)|^2$ which are shown in Fig.~\ref{fig:timeevolution}. Since the
  tunneling time $t^\uparrow$ gives the time for the first minimum 
 of $|U^\uparrow_{ii} (t)|^2$, it corresponds to the first maximum of $P_d$. At this minimum, 
  $|U^\uparrow_{ii} (t)|^2\simeq 0$ and 
 the value of $P_d$ at the first maximum is therefore very close to  $1/4$, which is also the long time uncorrelated value. 

We now turn to $P_d$ for finite temperature also plotted in Fig.~\ref{fig:Pd2}. We have chosen a slightly anisotropic Heisenberg system with  $\Delta=0.01$, which forces the staggered magnetization to be along the z-axis. The AF correlations are clearly visible as increased oscillations of $P_d$ as a function of time
 which is the main result of the present paper. In particular, note that the probability $P_d$ is \emph{larger} than $1/4$ for certain times which is a
 clear sign of AF ordering. To analyze this effect further, we consider the  high temperature regime $J/T \ll 1$ with $k_b = 1$, where
  \begin{equation}
 \left\langle \hat s^z_i \hat s^z_j \right\rangle=\frac{1}{4}\delta_{i,j}-\frac{J}{16 T}\delta_{\langle i,j\rangle}+{\mathcal{O}}(J^2/T^2)
\label{HighT}
 \end{equation}
  with $\delta_{\langle i,j\rangle}$  the Kronecker delta function connecting neighbor sites.
  Using this in Eq. (\ref{Pdsimple}) we get
 \begin{equation}
	P_d(i) = \frac 1 4(1- | U^\uparrow_{ ii} (t) |^2)+\frac{J}{8 T}  | U^\uparrow_{ \langle ii\pm 1\rangle} (t) |^2
\label{PdHighT}
\end{equation}
 for $J/T\ll 1$. Equation (\ref{PdHighT})
 clearly illustrates how AF correlations can increase $P_d$; for example as seen from Fig.~\ref{fig:timeevolution} at $t=t^\uparrow$, we have that $|U^\uparrow_{ii}(t^\uparrow)|^2 \simeq 0$ and $|U^\uparrow_{ii\pm 1}(t^\uparrow)|^2 \simeq 1/4$ such that $P_d(i) = \frac 1 4 (1+\frac{J}{8 T})$.
 As the temperature is lowered, the oscillations of $P_d$ becomes 
larger reflecting the increasing  AF correlations. 

\begin{figure}%[bt]
\includegraphics[clip=true,width=1\columnwidth]{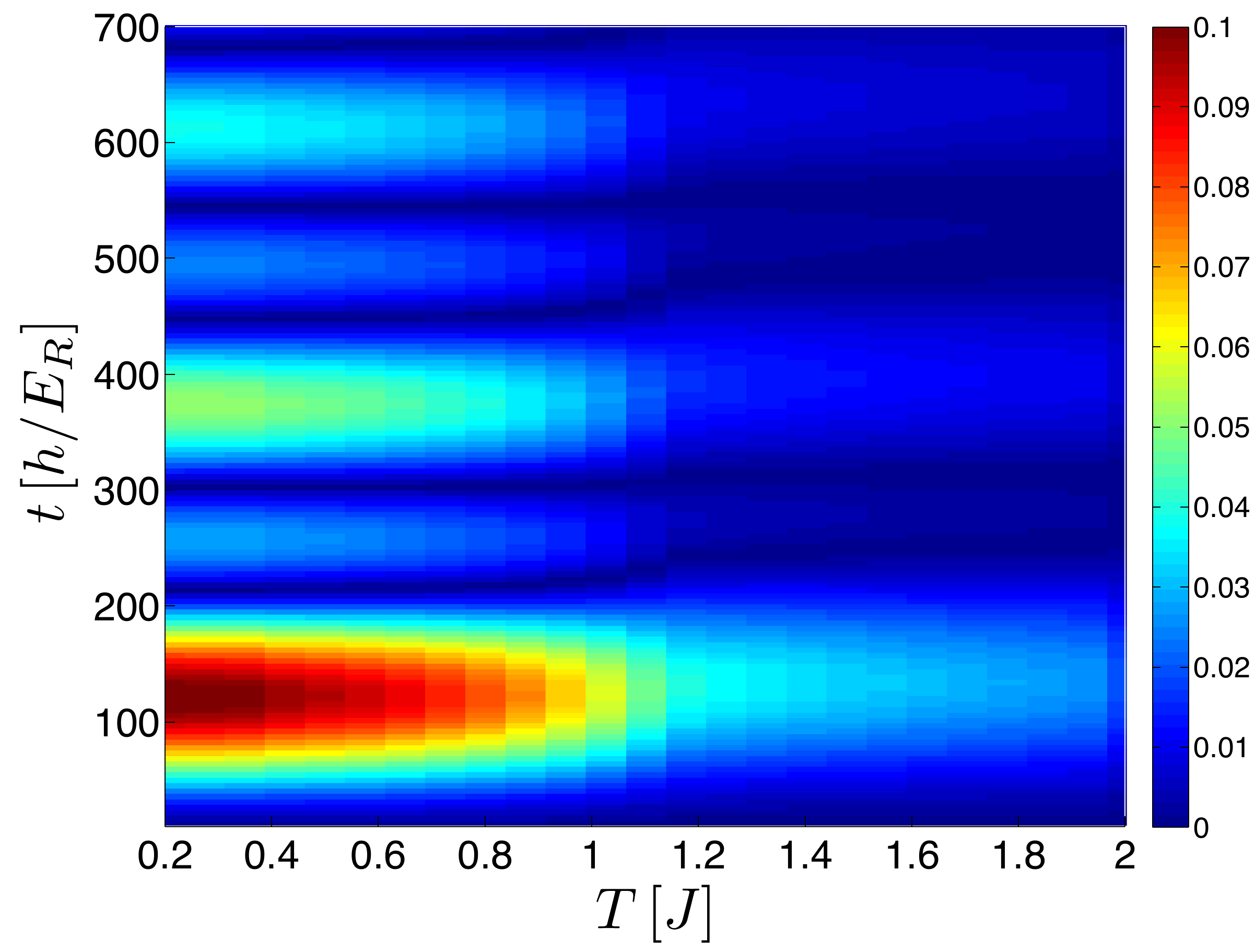}
\caption{(Color online) The difference in the double occupancy probability $\Delta P_d$ between the interacting and non-interacting cases as a function of time $t$ and temperature $T$.
}
\label{fig:Pd}
\end{figure}

We note  that the oscillations of $P_d$ increase significantly when 
$T<T_N$ where the system is in the broken symmetry phase. This is particularly visible for the later times which indicates 
the onset of long-range order. This effect is illustrated further in Fig.~\ref{fig:Pd}, where we plot the difference $\Delta P_d$ between the 
probabilities for finite and infinite temperatures. 
%%%%%%%%%%%%%%%%%%%%%%%%%%%%%%%%%%%%%%%%%%%%%%%%%%%%%%%%%%%%%%%%%%%
%%%%%%%%%%%%%%%%%%%%%%%%%%%%%%%%%%%%%%%%%%%%%%%%%%%%%%%%%%%%%%%%%%%
 %%%%%%%%%%%%%%%%%%%%%%%%%%%%%%%%%%%%%%%%%%%%%%%%%%%%%%%%%%%%%%%%%%%
\begin{figure}
\includegraphics[clip=true,width=.99\columnwidth]{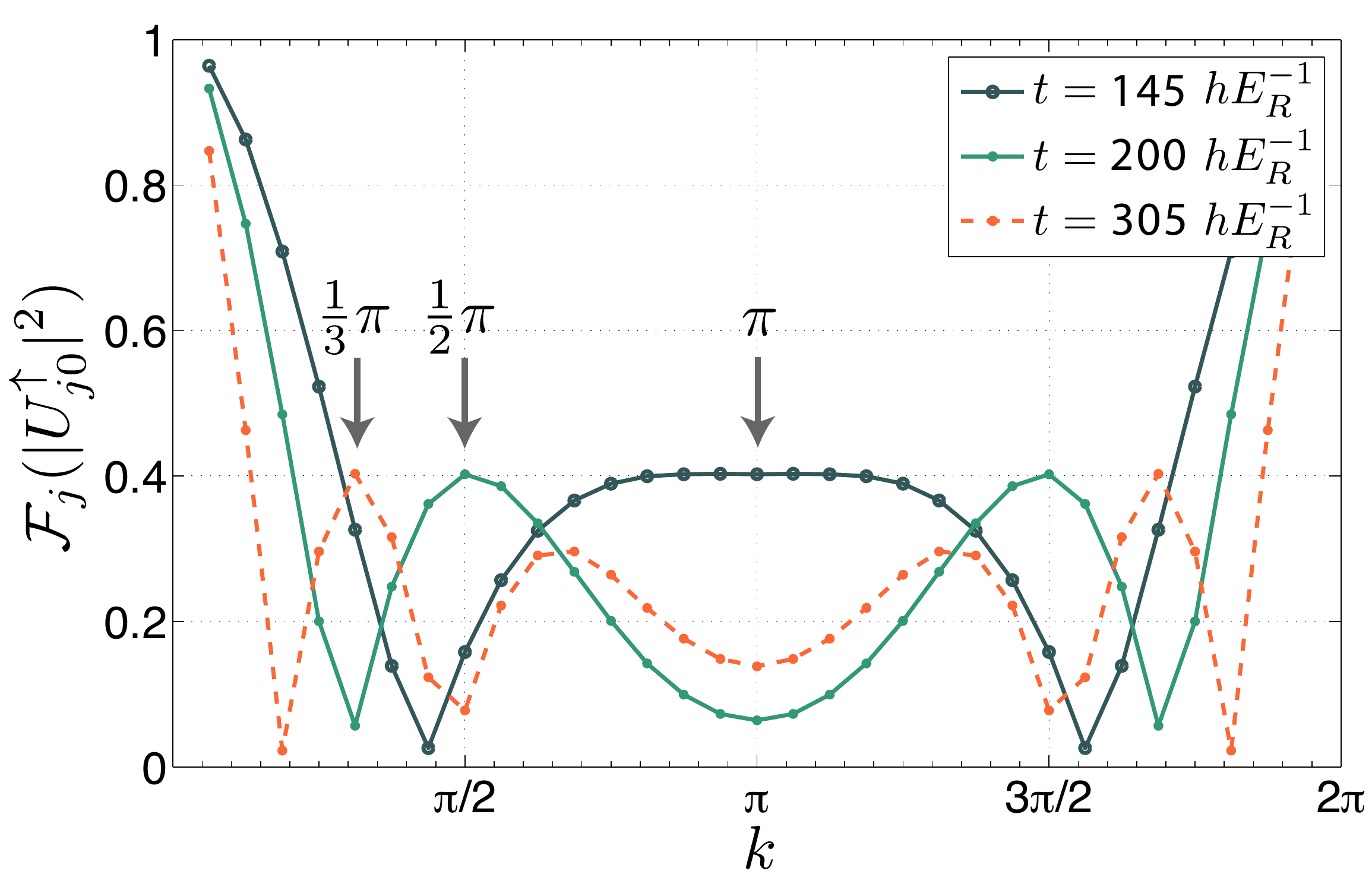}
\caption{(Color online) The $k$-space dependence of the norm square of the transfer matrix elements $U^{(2)}_{0j}$ for the first excited state shown at selected times. Note how the timed tunneling dynamics allows to probe the spin structures other than $k=\pi$, which is relevant for AF ordering, like the two examples shown here which mainly probe $k=\pi/2$ and $k=\pi/3$ ordered states.}
\label{fig:ft}
\end{figure} 
%%%%%%%%%%%%%%%%%%%%%%%%%%%%%%%%%%%%%%%%%%%%%%%%%%%%%%%%%%%%%%%%%%

We see that for the oscillations at later times, the difference increases significantly for $T<T_N\approx 0.945 J$ when long-range order sets in,
although the absolute size of the difference is rather small. Note also the two dips at $t \approx 260 \hbar / E_R$ and $t \approx 500 \hbar / E_R$ which shows up as long-range correlations set in at $T = T_N$. From Fig.~\ref{fig:Pd} one can also see that short-range correlations leave a signature at $T>T_N$, for example yielding an AF signal at $t\approx t^\uparrow$ of $\Delta P_d (T=2J) = 0.0195 \approx 0.08 P (T= \infty)$.

\section{Conclusions and Discussions}

In summary, we have demonstrated how to detect antiferromagnetic spin correlations in optical lattices by utilizing the difference in the tunneling dynamics of the two lowest lying vibrational states of the atomic wave-functions. Measurement of the number of doubly occupied sites yields a macroscopic signature of the magnetic correlations which is proportional to the number of lattice sites. Below the critical temperature $T_N$ the method has large signatures $\Delta P_d \approx 0.1$ as a result of the spin correlations. 

The method presented here is particularly relevant for measuring short range correlations as it provides an easily accessible experimental method, which requires a minimal change of the experimental setups already present. For measuring long range correlations the method is less suited as the spreading of the wave function for long times introduces an uncertainty in the correlation function measured. Such long correlation can thus be measured more precisely with different schemes, for instance relying on superlattice potentials~\cite{pedersen}.

Here we have mainly focussed on the detection of antiferromagnetic correlations, but the method may also be used to probe other spin ordered states. In Fig.~\ref{fig:ft} we show the $k$-space representation of $|U_{j0}^\uparrow|^2$ for three different selected times. As seen, by selecting appropriate times a magnetic ordering with periodicities of, for example, $k=\pi/2$ and $k=\pi/3$, can also be probed.

\section{Acknowledgements}

B.M.A. acknowledges support from The Danish Council for Independent Research $|$ Natural Sciences. O.F.S. acknowledges use of NOTUR computing facilities. A.S.S. acknowledges support from the Danish National Research Foundation.

\end{document}